\documentclass[fleqn,usenatbib]{mnras}

\usepackage{cuted}
\newenvironment{widetext}{\begin{strip}}{\end{strip}}

\usepackage{graphicx}
\usepackage{dcolumn}
\usepackage{bm}

\usepackage{float}
\usepackage{hyperref}
\hypersetup{colorlinks=true}
\usepackage{tabularx}
\usepackage{amssymb,amsmath,bm,tensor,braket}
\usepackage{xcolor}
\usepackage{enumerate}
\usepackage{mathtools}
\usepackage[capitalize]{cleveref}
\usepackage[utf8]{inputenc}
\usepackage[normalem]{ulem}
\usepackage[T1]{fontenc}
\usepackage[version=4]{mhchem}

\usepackage{natbib}

\usepackage{extarrows}
\usepackage{newtxtext,newtxmath}
\usepackage{orcidlink}

\newcommand{\dd}[0]{\mathrm{d}}

\allowdisplaybreaks

\begin{document}

\title[Tidal Deformation and Strain Accumulation of Solid Compact Stars]{Tidal Deformation and Strain Accumulation of Solid Compact Stars}
\author[H. Shen et al.]{
Hongxiang Shen,$^{1}$\thanks{shenhongxiang15057961620@stu.pku.edu.cn}
Yong Gao,$^{2}$\thanks{Corresponding author: yong.gao@aei.mpg.de}\,\orcidlink{0000-0003-1390-5477}
Hong-Bo Li,$^{3}$\thanks{Corresponding author: ihb2020@pku.edu.cn}\,\orcidlink{0000-0002-4850-8351}
Ren-Xin Xu$^{4,5,3}$\thanks{Corresponding author: r.x.xu@pku.edu.cn}\,\orcidlink{0000-0002-9042-3044}
\\
$^{1}$School of Physics, Peking University, Beijing 100871, China\\
$^{2}$Max Planck Institute for Gravitational Physics (Albert Einstein Institute), 14476 Potsdam, Germany\\
$^{3}$Kavli Institute for Astronomy and Astrophysics, Peking University, Beijing 100871, China\\
$^{4}$State Key Laboratory of Nuclear Physics and Technology, Peking University, Beijing 100871, China\\
$^{5}$Department of Astronomy, School of Physics, Peking University, Beijing 100871, China
}

\date{}

\maketitle

\begin{abstract}
The tidal deformability of compact stars encodes the equation of state of dense matter,
and gravitational-wave observations such as GW170817 have begun to constrain it
under the assumption of a fluid interior.
Yet whether the interior of pulsar-like compact stars is fluid or solid remains largely untested,
despite the distinct tidal responses the two states predict.
In this work, based on the
strangeon-star model, we develop a framework for modeling tidal deformation in
solid compact stars. Adopting a shear modulus of
$\mu = 10^{34}\,\mathrm{erg}\,\mathrm{cm}^{-3}$, we find a relative difference of
approximately $40\%$ in tidal deformability between solid and fluid strangeon
stars of $1.4\,M_\odot$, corresponding to a $\sim 10\%$ deviation from the
universal I--Love relation. We further model the accumulation of internal strain
during binary inspiral and find that it peaks near the stellar center. When the
gravitational-wave frequency reaches several hundred $\rm Hz$, large-scale
fracturing occurs and can release up to $\sim 10^{46}\,\mathrm{erg}$ of elastic
energy, sufficient to power short $\gamma$-ray-burst precursors. This
solid-to-fluid transition alters the tidal response and imprints on the waveform
and phase of the emitted gravitational radiation. Combined with the precursor
electromagnetic emission, these gravitational-wave signatures offer a
multi-messenger avenue to test the solid nature of pulsar-like compact
stars.

\end{abstract}

\begin{keywords}
dense matter -- equation of state -- gravitational waves -- stars: interiors -- gamma-ray burst: general
\end{keywords}

\section{Introduction}
\label{sec:intro}
The equation of state (EOS) of matter at super-nuclear densities remains an open
question. Because quantum chromodynamics is non-perturbative in this
regime, the EOS cannot be computed from first
principles~\citep{Kojo:2019raj}, and many competing models have been proposed for
the internal composition of pulsar-like compact stars. One influential
possibility originates with \citet{Witten:1984rs}, who conjectured that strange
quark matter—composed of nearly equal numbers of $u$, $d$, and $s$ quarks—could
be more stable than ordinary nuclear matter such as $^{56}$Fe, and might
therefore be the true ground state of strongly-interacting matter. Compact stars
made of such matter are known as strange quark stars (SqSs). At the densities
expected inside an SqS, quark matter may enter a crystalline
color-superconducting (CCS) phase~\citep{Alford:2000ze}, a supersolid whose shear
modulus is roughly 20–1000 times that of a neutron-star
crust~\citep{Mannarelli:2007bs}.

This CCS picture, however, is not the only one. \citet{Xu:2008nd} argued that the
density inside pulsar-like compact stars may be only a few times nuclear
density, where the color coupling remains strong and the CCS phase is not
guaranteed. In this regime quarks tend to condense in position space into clusters
that preserve three-light-flavor symmetry~\citep{Lai:2022yky,XU:2024ycj}, termed
strangeons, and a star built from them is a strangeon star (SnS). Although an SnS
shares its quark-level composition with strange quark matter, the clustered state
can have different macroscopic properties~\citep{Lai:2017ney}: in
particular, an SnS is expected to be a classical
solid~\citep{Xu:2003qf,Lai:2009cn}, in contrast to the fluid interior of a neutron
star and the supersolid interior of an SqS~\citep{Lau:2018mae}.

Despite their differences, the SnS and SqS hypotheses share a decisive feature: a
pure, bare solid star—classical solid for the SnS~\citep{Xu:2003qf} and supersolid
for the SqS~\citep{Lau:2017qtz}—with a large shear modulus of $\sim
10^{34}\,\mathrm{erg}\,\mathrm{cm}^{-3}$~\citep{Xu:2003xe,Mannarelli:2007bs}. This
is exactly what sets such stars apart from fluid neutron stars, which at most
carry a thin solid crust, and it leaves two observable imprints. First, a rigid
interior stiffens the star's tidal response, so a precise measurement of the
tidal deformability $\Lambda$ could tell a solid star from a fluid one. Second,
such a star can store substantial elastic energy and release it in starquakes—a
mechanism invoked to explain giant flares~\citep{Franco:1999pma,Xu:2003xe} and the
precursors of short $\gamma$-ray bursts (GRBs)~\citep{Zhou:2023dcf}. In this work we
quantify both signatures for solid strangeon stars: how the solid nature reshapes
the tidal response, and how much elastic energy can be released during a binary
inspiral.

We take the two signatures in turn. Tidal deformation encodes the internal
structure and EOS of a star, and has become directly observable with the rise of
gravitational-wave (GW) astronomy: the GW signal from a binary neutron-star
merger differs subtly from the binary-black-hole case because matter is
present~\citep{Flanagan:2007ix}, and this difference is quantified by the tidal
deformability~\citep{Damour:2009vw,Postnikov:2010yn,Hinderer:2009ca}. 
The detection of GW170817~\citep{LIGOScientific:2017vwq}, together
with subsequent
analyses,
placed a $90\%$ credible upper bound of $\Lambda \leq 800$ on a $1.4\,M_{\odot}$
neutron star, directly constraining the internal structure and EOS of compact
stars~\citep{De:2018uhw,LIGOScientific:2018cki,Annala:2017llu}.

Whether such a measurement can distinguish a solid star from a fluid one depends
on how strongly the shear modulus alters $\Lambda$. The relativistic theory of
tidal deformation, first developed by \citet{Hinderer:2007mb} and
\citet{Damour:2009vw}, shows that for a neutron star with only a thin solid crust
the correction to $\Lambda$ is negligible ($\sim
10^{-7}$)~\citep{Gittins:2020mll}; extensive calculations across many EOSs confirm
that the deviation from the universal I--Love relation is unobservable for GW
detection~\citep{Lin:2013nza,Biswas:2019ifs,Penner:2011pd,Alford:2007xm}, and even
a fluid strangeon star follows nearly the same relation as an ordinary neutron
star~\citep{Gao:2021uus}. A pure solid star is qualitatively different. For solid
SqSs, \citet{Lau:2017qtz} found a $\sim 60\%$ deviation, and we
show below (Section\,\ref{subsec:Tidal}) that solid strangeon stars exhibit a
$\sim 40\%$ deviation—large enough to be within reach of next-generation GW
detectors~\citep{Shterenberg:2024tmo}.

The second signature is the storage and release of elastic energy in starquakes.
\citet{1969Natur.223..597R} first studied quakes in the solid crust of a neutron
star and linked them to the sudden period changes known as glitches, and later
work associated the elastic energy released in such events with anomalous X-ray
pulsars and soft $\gamma$-ray
repeaters~\citep{Franco:1999pma,Ruderman_1991,Perna:2011gt}, a mechanism that may
also operate in SnSs~\citep{Xu:2003xe,Li:2023tng}. Because an SnS is solid
throughout rather than only in a thin crust~\citep{Xu:2003qf}, it can store far
more elastic energy—up to $\sim 10^{46}\,\mathrm{erg}$ even with $\sim 10^{-4}$ anisotropy
in pressure, and thus power more energetic
events, such as magnetar giant flares~\citep{Chen:2023pew} and GRB precursors. Building on
this, \citet{Zhou:2023dcf} proposed that the precursor of a short GRB
could originate from a tidally induced giant quake in a solid SnS, a
scenario they examined quantitatively for GRB 211211A~\citep{Xiao:2022quv}, whose
precursor is observationally confirmed. Competing explanations for such
precursors include resonant shattering~\citep{Suvorov:2020tmk,Suvorov:2022ldw,Kuan:2021jmk,Kuan:2021sin} and multiple-subjet
models~\citep{Nakamura:2000jn}; the distinctive claim here is that, for a pure
solid star, the stored elastic energy alone can suffice.

How much energy is released depends sensitively on the tidal field at the moment
of fracturing, so the crucial question is at what inspiral stage large-scale
fracturing sets in. \citet{Lai:2018ugk} argued that a solid SnS fractures on a
large scale once the tidal field drives the internal strain past a critical
threshold $\sigma$; taking $\mu = 4 \times 10^{32}\,\mathrm{erg}\,\mathrm{cm}^{-3}$
and $\sigma = 0.01$, they found this to occur at a GW frequency of $\sim
20\,\rm Hz$. Such large-scale fracturing would drive the star from a solid to a
fluid-like state, altering its tidal response and imprinting on the GW waveform
and phase~\citep{Lai:2018ugk}. Both the transition frequency and the size of the
phase shift depend on $\mu$ and $\sigma$, so the GW signal alone can constrain
these two parameters.

The paper is organized as follows. Section\,\ref{sec:Meth} introduces the
modeling and theoretical framework for perturbations inside the star, deriving
the differential equations (Section\,\ref{subsec:Perturb}) and the corresponding
boundary conditions (Section\,\ref{subsec:bound}). This section is rather
technical; readers primarily interested in the main results may proceed directly
to Section\,\ref{sec:Res}. We then present numerical results, including the
tidal deformability (Section\,\ref{subsec:Tidal}) and strain accumulation
(Section\,\ref{subsec:strain}). Discussion and conclusions are provided in the
final section. Unless otherwise noted, geometric units with $G=c=1$ are used
throughout this work, where $G$ is the gravitational constant and $c$ is the speed of light.

\section{Methodology}
\label{sec:Meth}

In this section, we present the equation of state, background solutions, and
perturbation equations.

\subsection{Equation of state}
\label{subsec:EOS}

Our treatment is based on the EOS for strangeon
matter~\citep{Zhang:2023mzb}, in which the interaction between quark clusters is
described by a Lennard-Jones potential~\citep{Jones_1924},
\begin{equation}
U(r)=4\epsilon \left( \left(\frac{r_0}{r}\right)^{12} - \left(\frac{r_0}{r}\right)^6 \right)\, ,
\end{equation}
where $r$ is the inter-cluster distance and $\epsilon$ is the depth of the
potential well. Summing this interaction over the lattice yields the energy
density $\rho$ and pressure $P$ as functions of the cluster number density $n$,
\begin{align}
\rho &= 2\epsilon \left( A_{12} r_0^{12} n^5 - A_6 r_0^6 n^3 \right) + n N_q m_q\, , \\
P &= n^2 \frac{\dd\,(\rho/n)}{\dd n} = 4\epsilon \left( 2A_{12} r_0^{12} n^5 - A_6 r_0^6 n^3 \right)\, ,
\end{align}
where the lattice constants $A_{12}=6.2$ and $A_6=8.4$, the well depth
$\epsilon = 25\,\mathrm{MeV}$, and the quark mass $m_q = 300\,\mathrm{MeV}$ are
adopted from~\citep{Zhang:2023mzb,Gao:2021uus}. Here $N_q$ is the number of
quarks in a single cluster, taken as $N_q=18$~\citep{Zhou:2017pha}, so that the
term $n N_q m_q$ accounts for the rest-mass energy.

The EOS is fixed once the surface density is specified. Requiring $P = 0$ at the
stellar surface gives the surface cluster number density $\sqrt{A_6/(2 A_{12}
r_0^6)}$, which corresponds to a baryon number density
\begin{equation}
\label{eq:EOSns}
    n_s = \sqrt{\frac{A_6}{2 A_{12}}}\frac{N_q}{3 r_0^3}\,.
\end{equation}
We adopt $n_s = 0.36\,\rm fm^{-3}$, more than double the value
$0.16\,\rm fm^{-3}$ of typical nuclear matter. Equation~(\ref{eq:EOSns}) then
determines $r_0$, and the EOS is fully specified.

\subsection{Background}
\label{subsec:background}
For a spherically symmetric, static stellar configuration, the metric can be
written as:
\begin{equation}
\dd s^2 = -e^{\nu(r)} \dd t^2 + e^{\lambda(r)} \dd r^2 + r^2 (\dd \theta^2 + \sin^2\theta \dd\phi^2)\, .
\end{equation}

\begin{figure}
    \centering
    \includegraphics[width = 0.48\textwidth]{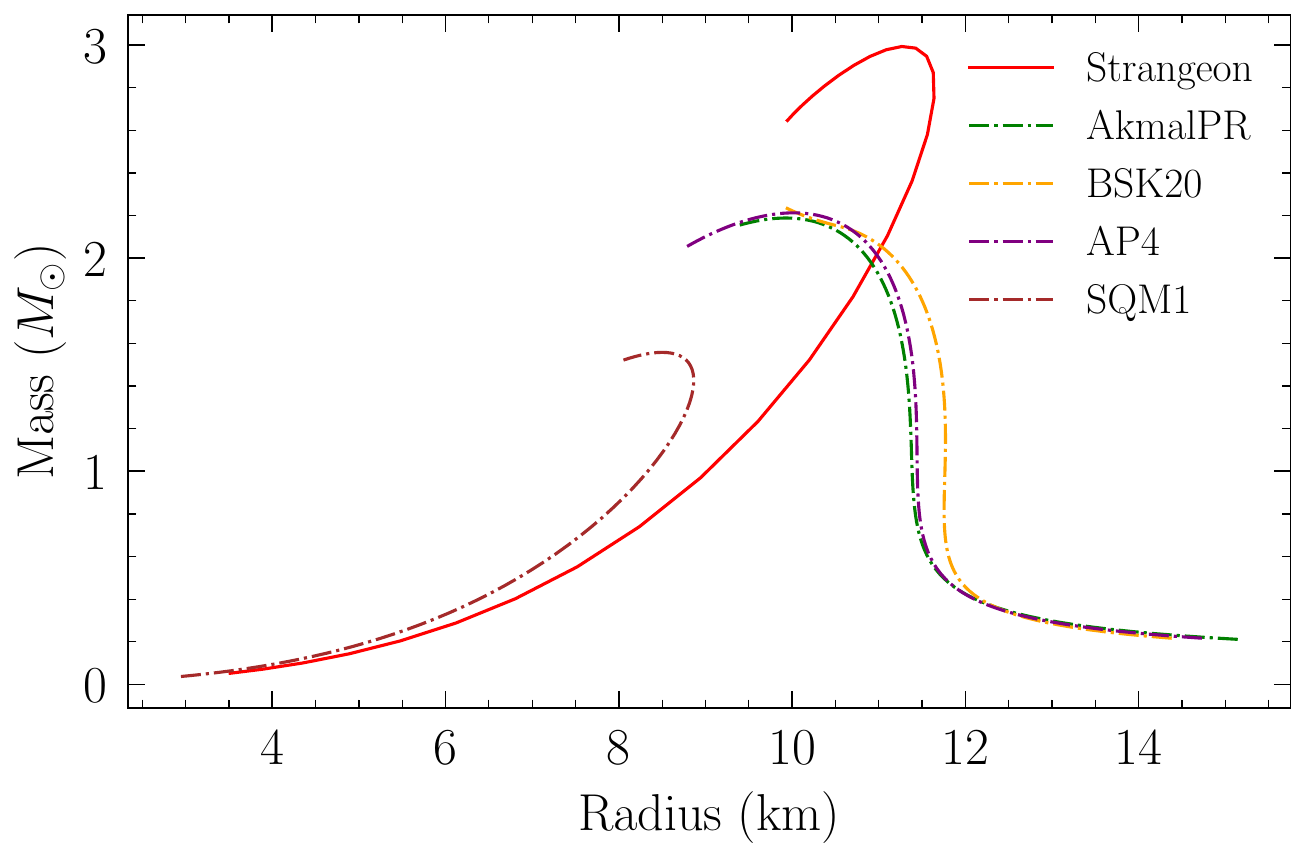}
    \caption{Mass--radius relation for compact stars. The red curve corresponds to the strangeon matter EOS with the parameters described in Section\,\ref{subsec:EOS}. The green curves represent several representative neutron star EOSs for comparison.}
    \label{fig:figure2}
\end{figure}

The Tolman--Oppenheimer--Volkoff equations then follow:
\begin{align}
\frac{\dd \nu}{\dd r} &= \frac{2e^{\lambda(r)}}{r^2}\left(m(r) + 4\pi r^3 P(r)\right)\, , \\
\frac{\dd P}{\dd r} &= -\frac{\rho(r) + P(r)}{2} \nu'(r)\, , \\
\frac{\dd m}{\dd r} &= 4\pi r^2 \rho(r)\, .
\end{align}
The system is supplemented by the relation
\begin{equation}
e^{\lambda(r)} = \frac{1}{1 - 2m(r)/r}\, ,
\end{equation}
where $m(r)$ is the mass enclosed within radius $r$. Integrating this system for
a given central energy density yields the radial profiles of the pressure $P(r)$
and the metric functions $\nu(r)$ and $\lambda(r)$, and fixes the stellar mass
and radius. The resulting mass--radius relation is shown in
Fig.\,\ref{fig:figure2}.

\subsection{Perturbation equations}
\label{subsec:Perturb}
We begin by establishing a spherical coordinate system with the polar axis
aligned toward the companion star. Given the rotational symmetry of the tidal
field around this axis, both the displacement vector $\xi^a$ and the metric
perturbation $\delta g_{ab}$ in the Regge--Wheeler metric~\citep{Regge:1957td}
can be expanded in terms of spherical harmonics as follows~\citep{Thorne_1967}:

\begin{align}
    \xi^r &= \frac{W(r)}{r}Y_{lm}(\theta,\phi)\, ,\\
    \xi^\theta &= \frac{V(r)}{r^2}\partial_\theta Y_{lm}(\theta,\phi)\, ,\\
    \xi^\phi &= \frac{V(r)}{r^2\sin^2\theta}\partial_\phi Y_{lm}(\theta,\phi)\, ,
\end{align}

\begin{equation}
\delta g_{ab}=h_{ab}(r)Y_{lm}(\theta,\phi)e^{i\omega t} \,,
\end{equation}
where
\begin{equation}
h_{ab}(r)=
    \begin{pmatrix}
H_0(r)e^{\nu} & i\omega H_1(r) & 0 & 0 \\
i\omega H_1(r) & H_2(r)e^{\lambda} & 0 & 0 \\
0 & 0 & r^2K(r) & 0 \\
0 & 0 & 0 & r^2\sin^2\theta K(r)
    \end{pmatrix} \,.
\end{equation}

For a solid star undergoing elastic deformation, the perturbation of the
energy-momentum tensor can be expressed as the sum of two components:
\begin{equation}
\delta T_{ab} = \delta T^{\mathrm{bulk}}_{ab} + \delta T^{\mathrm{shear}}_{ab}\,,
\end{equation}
where $\delta T_{ab}^{\mathrm{bulk}}$ accounts for changes in the
energy-momentum tensor due to the deformation itself and is given by
\begin{equation}
\delta T^{\mathrm{bulk}}_{ab} = (\delta\rho + \delta P) U_a U_b + \delta P g_{ab} + (P + \rho)(U_a \delta U_b + \delta U_a U_b).
\end{equation}

The term $\delta T_{ab}^{\mathrm{shear}}=-2\mu \delta S_{ab}$ describes the
contribution from elastic stresses, where $\delta S_{ab}$ is the strain tensor.
This tensor can also be expanded in spherical harmonics as follows:
\begin{equation}
\begin{split}
\delta & S^r_r = S_r Y_{lm}(\theta, \phi) \\
&= \frac{1}{3}\left(H_2 - K + \frac{l(l+1)}{r^2}V + \frac{2}{r}\frac{\dd W}{\dd r} - \left(\frac{4}{r^2} - \frac{\lambda'}{r}\right)W \right) Y_{lm}\, ,
\end{split}
\end{equation}
\begin{equation}
\begin{split}
\frac{1}{r}& \delta S^r_A = S_{\perp} \partial_{A} Y_{lm}(\theta, \phi) \\
&= \frac{e^{-\lambda}}{2r} \left( \frac{\dd V}{\dd r} - \frac{2V}{r} + e^{\lambda} \frac{W}{r} \right) \partial_A Y_{lm}\, ,
\end{split}
\end{equation}
where $A$ denotes the angular coordinates $\theta$ and $\phi$.

To characterize the strain magnitude, we introduce the parameters $Z_r$ and
$Z_{\perp}$, defined as:
\begin{align}
Z_r(r) &= \Delta P(r) - 2\mu S_r(r)\, , \\
Z_\perp(r) &= -2\mu S_\perp(r)\, .
\end{align}
Furthermore, we define an auxiliary parameter $J$ as:
\begin{equation}
J = H_0' - 8\pi e^{\lambda} (\rho + P) \frac{W}{r} + 16\pi \nu' \mu V\, .
\end{equation}

Substituting these parameters into the Einstein field equations and taking the
static perturbation limit (i.e.\ $\omega=0$), we obtain a system of six
first-order linear ordinary differential equations for the variables $(H_0, J,
W, V, Z_r, Z_\perp)$, as shown in Appendix \ref{app:A}. These equations,
together with the boundary conditions derived below, fully determine the tidal
deformability and the internal strain distribution of the star.

\subsection{Boundary conditions}
\label{subsec:bound}
\subsubsection{Center}
\label{subsubsec:center}

As $r\rightarrow 0$, the differential equations
(\ref{eq:dif1}--\ref{eq:dif2}) contain divergent terms, which must be
regularized by imposing appropriate boundary conditions on the allowed behavior
at the center. Performing a Taylor expansion around
$r=0$, we assume the following leading-order forms:
\begin{align}
    \label{eq:initial}
    H_0 &= r^2\left(H_0^{(0)}+H_0^{(2)}r^2\right)\, ,\\
    J &= r\left(J^{(0)}+J^{(2)}r^2\right)\, ,\\
    W &= r^2\left(W^{(0)}+W^{(2)}r^2\right)\, ,\\
    V &= r^2\left(V^{(0)}+V^{(2)}r^2\right)\, ,\\
    Z_{\perp} &= Z_{\perp}^{(0)}+Z_{\perp}^{(2)}r^2\, ,\\
    Z_r &= Z_r^{(0)}+Z_r^{(2)}r^2\, .
    \label{eq:initial2}
\end{align}

Substituting these expansions into the differential equations and requiring that
the coefficients of the divergent terms vanish and that the leading non-zero
terms match order by order, we obtain three independent regular eigensolutions.
These can be expressed in terms of three free parameters:
$(H_0^{(0)},V^{(0)},V^{(2)})$. The relations for the remaining expansion
coefficients are derived accordingly, as shown in Appendix \ref{app:B}.

\subsubsection{Surface}
\label{subsubsec:surface}

At the stellar surface, integrating the differential equations reveals that the
quantities $(H_0,J,W,Z_{\perp},Z_r)$ are continuous across the
boundary~\citep{Penner:2011pd}. In vacuum the stress vanishes, and the auxiliary
parameter simplifies to $J=H_0'$. Consequently, the following boundary
conditions are obtained, where $R_+$ denotes the outer (vacuum) side, $R_-$ the
inner side, and $R$ the stellar radius:
\begin{align}
    Z_{\perp}&(R_+) =Z_{\perp}(R_-)=0\, ,\\
    Z_r&(R_+) =Z_r(R_-)=0\, ,\\
    H_0'&(R_+) =J(R_+)=J(R_-)\, ,\\
    H_0&(R_+) =H_0(R_-)\, .
\end{align}

Combined with the central boundary conditions specified in
Section\,\ref{subsubsec:center}, the solution to the differential equations is
determined up to an overall multiplicative factor. This factor depends on the
strength of the tidal field. However, for calculating the tidal
deformability—which depends only on the logarithmic derivative
$y=RH_0'(R_+)/H_0(R_+)$—the solution thus obtained is already
sufficient.

We have verified these boundary conditions and performed numerical calculations
that successfully reproduce the results presented by~\citet{Lau:2018mae}.

\subsubsection{Vacuum}
\label{subsubsec:Vacuum}

In vacuum, the perturbation equation for the metric admits an analytical solution:
\begin{equation}
    H_0''+(2/r-\lambda')H_0'-\left[\frac{l(l+1)e^{\lambda}}{r^2}-\lambda'^2\right]H_0=0\, .
\end{equation}
Introducing the variable $x=r/M - 1$, the solution can be expressed as a linear
combination of associated Legendre functions~\citep{Thorne_1967}:
\begin{equation}
\label{eq:the}
    H_0(r)=c_1Q_l^2(x)+c_2P_l^2(x)\, .
\end{equation}
At large distances from the origin, these functions exhibit the following
asymptotic behaviors:
\begin{align}
    Q^2_l(x)&\sim \left(\frac{M}{r}\right)^{l+1}\, ,\\
    P^2_l(x)&\sim \left(\frac{r}{M}\right)^{l}\, .
\end{align}

For the tidal field, the metric component can be expanded in powers of $r$
as~\citep{Thorne:1997kt}:
\begin{equation}
\label{eq:the2}
    -\frac{1+g_{tt}}{2}=-\frac{M}{r}-\frac{3Q_{ij}}{2r^3}\left(\frac{x^ix^j}{r^2}-\frac{1}{3}\delta^{ij}\right)+\frac{1}{2}\mathcal{E}_{ij}x^ix^j\, ,
\end{equation}
where $\mathcal{E}_{ij}$ characterizes the strength of the tidal field. The
tidal Love number $k_2$ for a static, spherically symmetric star is defined
by~\citep{Hinderer:2007mb}:
\begin{equation}
    \label{eq:k2}
    Q_{ij}=-\frac{2}{3}k_2 R^5 \mathcal{E}_{ij}\, .
\end{equation}

Comparing the asymptotic behavior of Eq.\,(\ref{eq:the}) with $l=2$ to the
expansion in Eq.\,(\ref{eq:the2}) as $r\rightarrow\infty$, the Love number is
found to be
\begin{equation}
    \label{eq:lovenumber}
    k_2 = \frac{4}{15}\left(\frac{M}{R}\right)^5\frac{c_1}{c_2}\, .
\end{equation}

Thus, we have the ``normalized tidal deformability'' $\Lambda$
\begin{equation}
    \Lambda = \frac{2}{3}k_2\left(\frac{R}{M}\right)^5\, .
\end{equation}

The coefficients $c_1$ and $c_2$ can be determined from the values of $H_0$ and
$H_0'$ at the stellar surface, which are obtained from the interior solution
(Section\,\ref{subsubsec:center}). Note that these values are defined up to an
overall multiplicative factor, as discussed in Section\,\ref{subsubsec:surface}.
Substituting $r=R$ into Eq.\,(\ref{eq:the}) and its derivative, we
obtain~\citep{Gittins:2021zpv}:
\begin{widetext}
\begin{equation}
    c_1=\frac{R(R-2M)}{8M^3}\left[R(2M-R)H_0'(R_+)+2(R-M)H_0(R_+)\right]\, ,
\end{equation}
\begin{equation}
\label{eq:c21}
\begin{split}
    c_2 ={}& \frac{1}{48M^3R(2M-R)}\Bigg\{
        -2M\Big[R\left(4M^4+6M^3R-22M^2R^2+15MR^3-3R^4\right)H_0'(R_+)\\
        &+2\left(M^4-2M^3R+13M^2R^2-12MR^3+3R^4\right)H_0(R_+)\Big]\\
        &+3R^2(R-2M)^2\left[R(2M-R)H_0'(R_+)+2(R-M)H_0(R_+)\right]\ln{\frac{R}{R-2M}}
        \Bigg\}\, .
\end{split}
\end{equation}
\end{widetext}

From the perspective of binary orbital dynamics, the gravitational-wave
frequency can be expressed as:
\begin{equation}
    f_{\mathrm{GW}}=\frac{1}{\pi}\sqrt{\frac{M+M_{\mathrm{comp}}}{d^3}}\, ,
\end{equation}
where $d$ is the orbital separation and $M_{\mathrm{comp}}$ is the mass of the
companion. The tidal field strength, for $l=2,m=0$, is given
by~\citep{Gittins:2020mll}:
\begin{equation}
    \mathcal{E}_{ij}=\frac{\partial^2\Phi_{\mathrm{ext}}}{\partial x^i\partial x^j}=-3\frac{M_{\mathrm{comp}}}{d^3}(n_in_j-\frac{1}{3}\delta_{ij})\, .
\end{equation}
Comparing the asymptotic form of Eq.\,(\ref{eq:the}) with the tidal expansion in
Eq.\,(\ref{eq:the2}), the coefficient $c_2$ is identified as
\begin{equation}
\label{eq:c22}
    c_2=\frac{2}{3}\sqrt{\frac{4\pi}{5}}\frac{M^2M_{\mathrm{comp}}}{d^3}=\frac{2\pi^2}{3}\sqrt{\frac{4\pi}{5}}\frac{M^2M_{\mathrm{comp}}}{M+M_{\mathrm{comp}}}f^2_{\mathrm{GW}}\, .
\end{equation}

Since the expression for $c_2$ derived from the vacuum solution,
Eq.\,(\ref{eq:c21}), must equal that obtained from the tidal field,
Eq.\,(\ref{eq:c22}), the remaining overall scaling factor is determined for a
given GW frequency. This completes the solution of the boundary value problem
for the system of ordinary differential equations.
\section{Results and discussion}
\label{sec:Res}

\subsection{Tidal deformation}
\label{subsec:Tidal}

\begin{figure}
    \centering
    \includegraphics[width=0.48\textwidth]{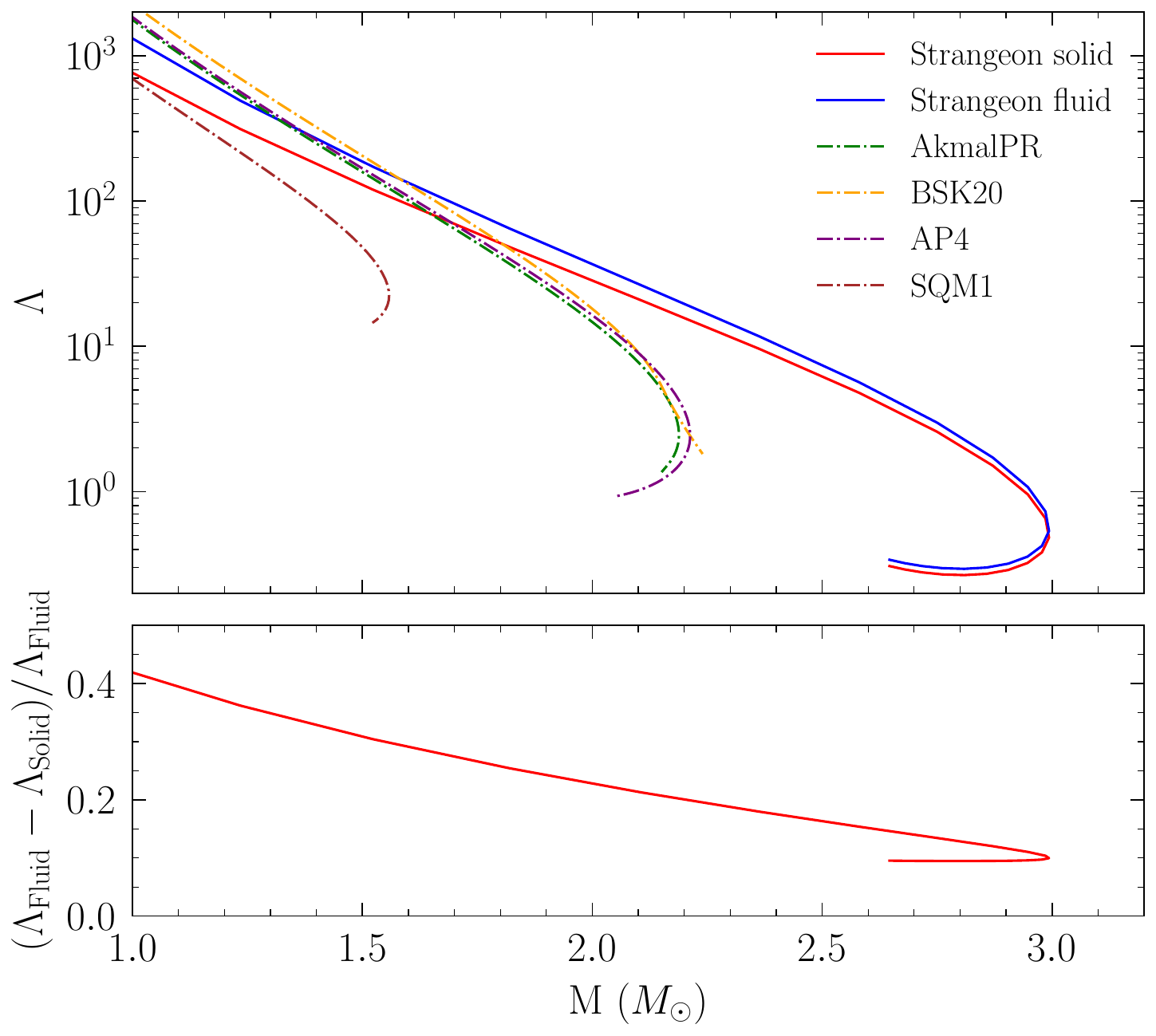}
    \caption{Tidal deformability $\Lambda$ as a function of stellar mass for strangeon stars, showing a $\sim 40\%$ relative difference at intermediate masses. The red curve corresponds to solid stars ($\mu = 10^{34}\,\mathrm{erg}\,\mathrm{cm}^{-3}$), and the blue curve to fluid stars. The relative difference $(\Lambda_{\rm fluid} - \Lambda_{\rm solid})/\Lambda_{\rm fluid}$ is also shown.}
    \label{fig:figure3}
\end{figure}

Employing the methodology outlined in Section\,\ref{sec:Meth}, we compute the
tidal deformabilities for both solid and fluid states, denoted $\Lambda_{\rm
solid}$ and $\Lambda_{\rm fluid}$, respectively. In the calculation of
$\Lambda_{\rm solid}$, a typical shear modulus value of $\mu=10^{34}\, \rm erg
\, cm^{-3}$ is adopted~\citep{Xu:2003xe, Zemlyakov:2025qqp, Mannarelli:2007bs}.
The results are presented in Fig.\,\ref{fig:figure3}. For SnSs of intermediate
masses, the relative difference in tidal deformability between the two states
reaches approximately $40\% $.

\begin{figure}
    \centering
    \includegraphics[width = 0.48\textwidth]{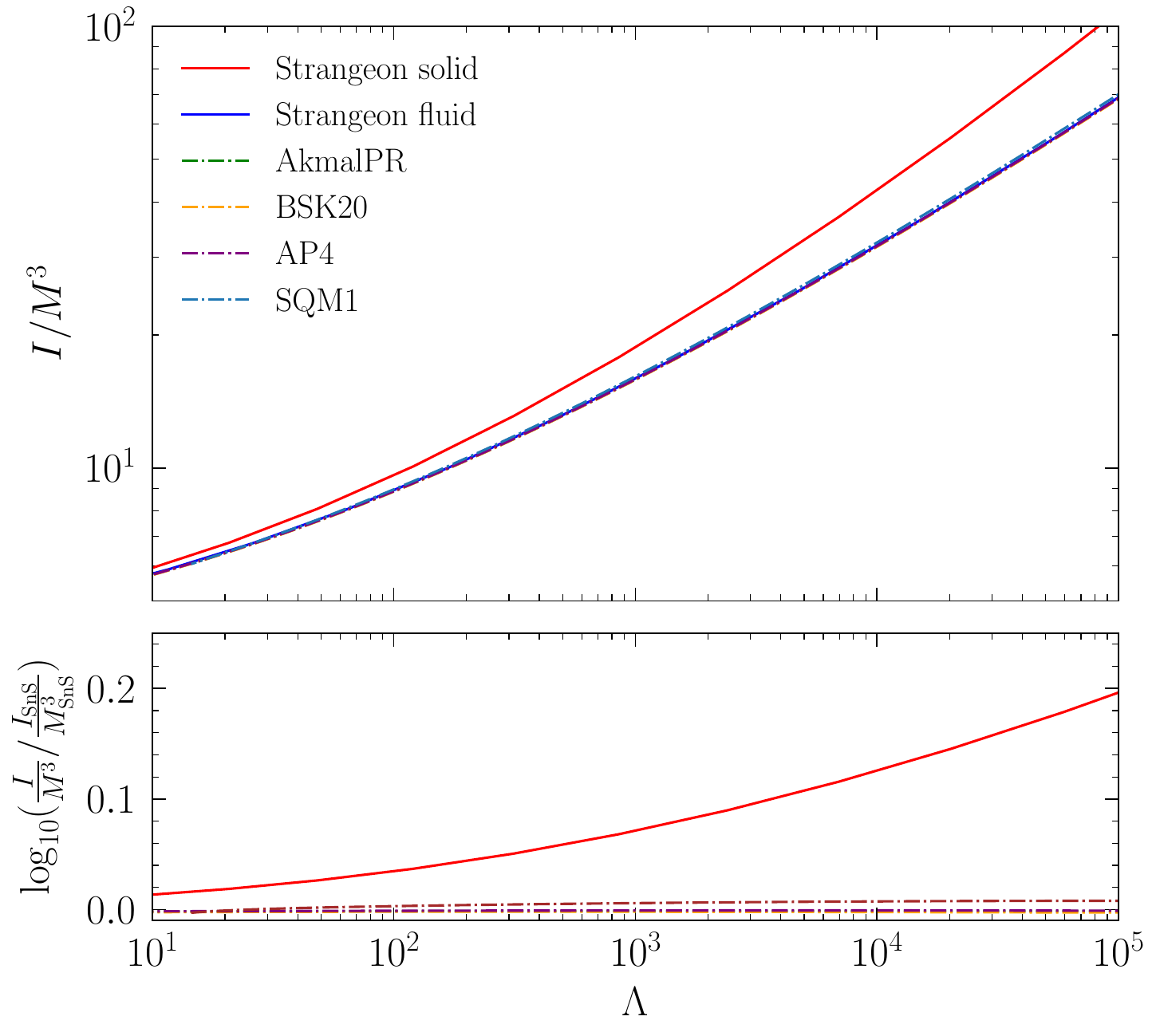}
    \caption{Deviation from the universal I--Love relation. Solid strangeon stars (red) exhibit a significant departure from the universal curve followed by fluid neutron stars and fluid strangeon stars (blue), demonstrating that the solid nature can be identified through combined measurements of the moment of inertia and tidal deformability.}
    \label{fig:I-Love}
\end{figure}

In contrast, calculations for neutron stars with a solid crust report a relative
difference of less than $10^{-7}$~\citep{Gittins:2020mll}. Meanwhile,
\citet{Gao:2021uus} showed that there is little difference in the universal
I--Love relation between a fluid strangeon star and a normal neutron star. Our
work presents a different picture, as shown in Fig.\,\ref{fig:I-Love}: by
accounting for the solid nature of strangeon stars, it becomes possible to
distinguish them from ordinary neutron stars with future, more sensitive GW
observations.

The screening effect has been investigated by \citet{Lau:2018mae} in the context
of ``dressed quark stars'', which consist of a CCS quark matter core ($\sim 9\,\rm
km$) with a thin nuclear matter crust ($\sim 0.21\,\rm km$). Their results show
that, compared with the bare solid star, the relative difference in tidal
deformability is less than $5\%$, regardless of whether the crust is solid or
fluid. This weak screening stems from the low energy density of the thin nuclear
layer relative to the core. Strangeon matter shares a similar quark-level
composition and likewise retains a non-zero energy density at its surface,
producing a sharp density discontinuity at any core-envelope interface. The
screening should therefore be equally weak for an SnS, so that even a thin
nuclear ocean or crust would leave the $\sim 40\%$ relative difference between
solid and fluid configurations largely intact.

\begin{figure}
    \centering
    \includegraphics[width=0.48\textwidth]{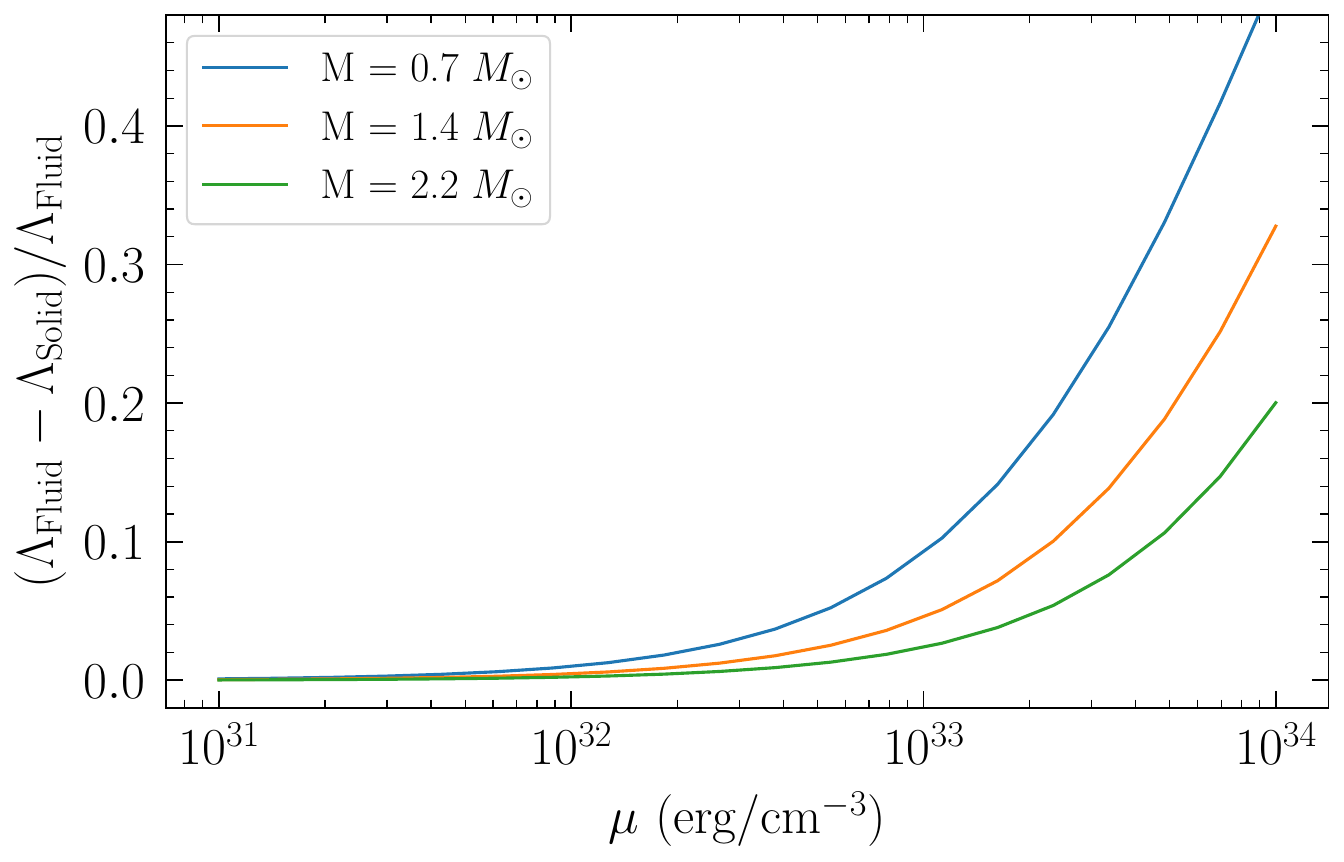}
    \caption{Relative difference in tidal deformability between solid and fluid strangeon stars as a function of shear modulus $\mu$, for three representative stellar masses. As expected, the difference vanishes in the limit $\mu \rightarrow 0$.}
    \label{fig:figure4}
\end{figure}

Furthermore, we examine how the relative difference in tidal deformability
varies with the shear modulus. Qualitatively, as $\mu\rightarrow 0$, the tidal
deformability of a solid star should converge to that of its fluid counterpart.
As shown in Fig.\,\ref{fig:figure4}, this expected behavior is consistently
confirmed for stars of different masses.

This behavior can be understood through a simple Newtonian estimate. Building on the classical definition of the Love number by \citet{Love1911}, \citet{Greff-Lefftz2005} and others derived a compact analytical relation between the Love number $k_2$, the shear modulus $\mu$, and the EOS of a solid star,
\begin{equation}
    \label{eq:k2newton}
    k_2 = \frac{3/4}{1+\tilde{\mu}}\,,
\end{equation}
where $\tilde{\mu} = 19\mu/(2\rho g R)$, $g$ is the surface gravitational acceleration, and $R$ is the stellar radius. For a constant-density, incompressible solid star, this relation is exact; we confirm this explicitly in Appendix\,\ref{app:C}. From Eq.\,(\ref{eq:k2newton}), the relative difference in tidal deformability between a solid and a fluid star (the latter corresponding to $\mu = 0$) is immediately obtained as:
\begin{equation}
   \frac{\Lambda_{\rm fluid}-\Lambda_{\rm solid}}{\Lambda_{\rm fluid}} = \frac{k_{2,\rm fluid}-k_{2,\rm solid}}{k_{2,\rm fluid}} = \frac{\tilde{\mu}}{1+\tilde{\mu}}\,.
\end{equation}

To compare with our numerical results, we evaluate $\tilde{\mu}$ for a typical SnS. In general relativity, the effective surface gravity is enhanced by a factor of $\left(1-2GM/(Rc^2)\right)^{-1/2}$ compared to the Newtonian value. Adopting the fiducial parameters $M = 1.4\,M_{\odot}$, $R = 9.96\,\rm km$, $\mu = 10^{34}\,\rm erg\,cm^{-3}$, together with the mean density $\rho = 3M/(4\pi R^3)$ and the relativistic surface gravity $g = \frac{GM}{R^2}\left(1-2GM/(Rc^2)\right)^{-1/2}$, we find $\tilde{\mu} = 0.58$. This yields:
\begin{equation}
   \frac{\Lambda_{\rm fluid}-\Lambda_{\rm solid}}{\Lambda_{\rm fluid}} = \frac{\tilde{\mu}}{1+\tilde{\mu}} = 0.37\,.
\end{equation}
This agrees closely with our full numerical result of approximately $40\%$. This consistency not only validates our numerical implementation but also demonstrates that the essential physics of the solid-fluid tidal deformability difference is already captured by the Newtonian estimate.

\subsection{Strain accumulation and fracturing}
\label{subsec:strain}

Beyond the tidal deformability, we must also quantify the strain that builds up
inside the star. As the binary inspirals, the orbital separation shrinks and the
GW frequency rises, strengthening the tidal field and driving progressively
larger strain into the stellar interior. We measure its magnitude by the squared
norm of the strain tensor~\citep{Gittins:2020mll, Gittins:2021zpv,Gao:2025aqo},
\begin{equation}
    s^2 = \frac{3}{2}\delta s_{ab}\delta s^{ab}\, .
\end{equation}

Substituting the definition of $\delta s_{ab}$ from
Section\,\ref{subsec:Perturb}, we obtain:
\begin{equation}
\label{eq:s2}
\begin{split}
    s^2 &= \frac{45}{256\pi r^4\mu^2}[(3\cos^2{\theta}-1)^2(\Delta P-Z_r)^2\\
    +&12e^{-\lambda}\sin^2(2\theta) Z_{\perp}^2+48\sin^4\theta \mu^2 V^2]\, .
\end{split}
\end{equation}

\begin{figure}
    \centering
    \includegraphics[width=0.48\textwidth]{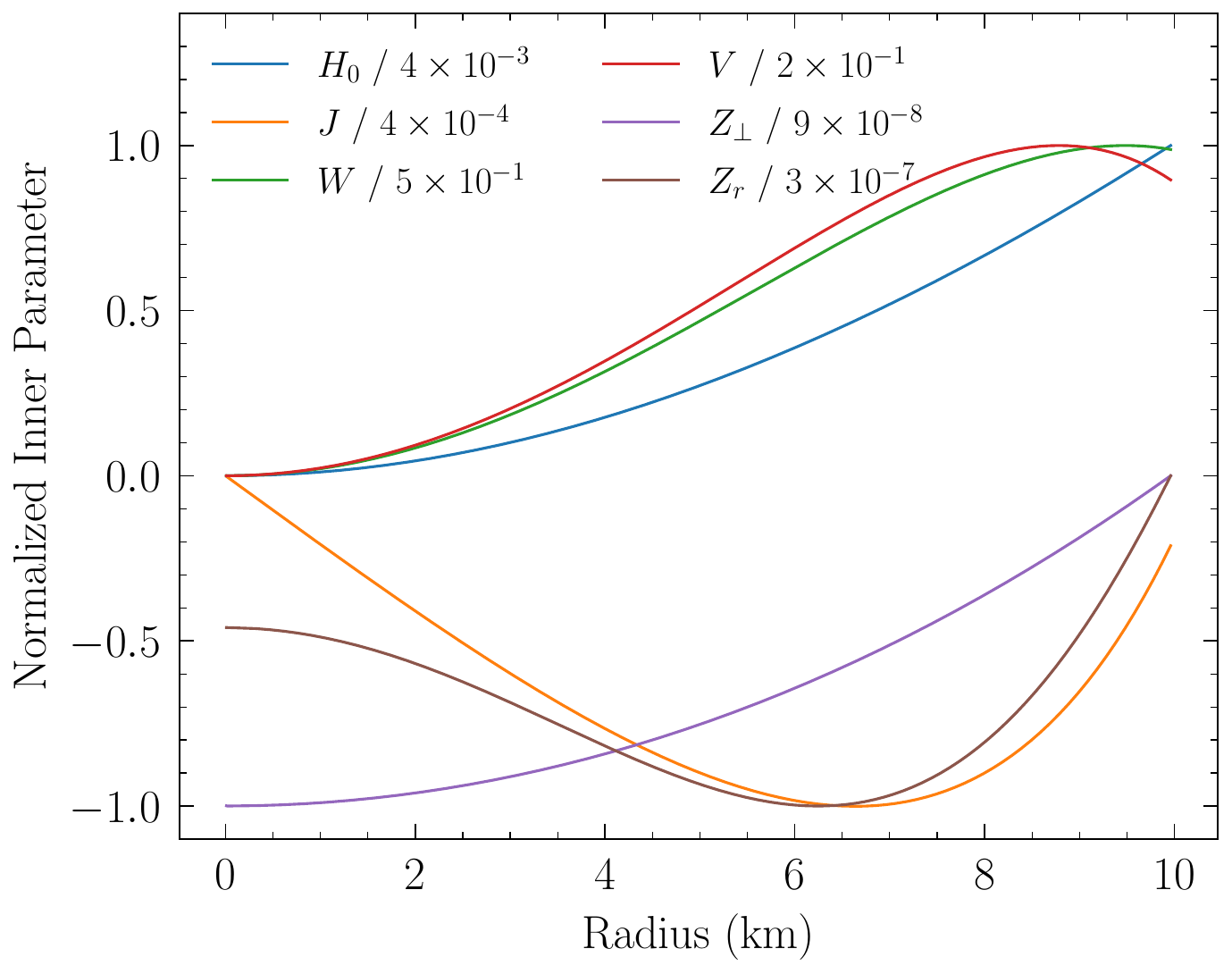}
    \caption{Radial profiles of the perturbation variables inside a $1.4 \,M_{\odot}$ strangeon star with shear modulus $\mu = 10^{34}\,\mathrm{erg}\,\mathrm{cm}^{-3}$.}
    \label{fig:figure5}
\end{figure}

\begin{figure}
    \centering
    \includegraphics[width = 0.48\textwidth]{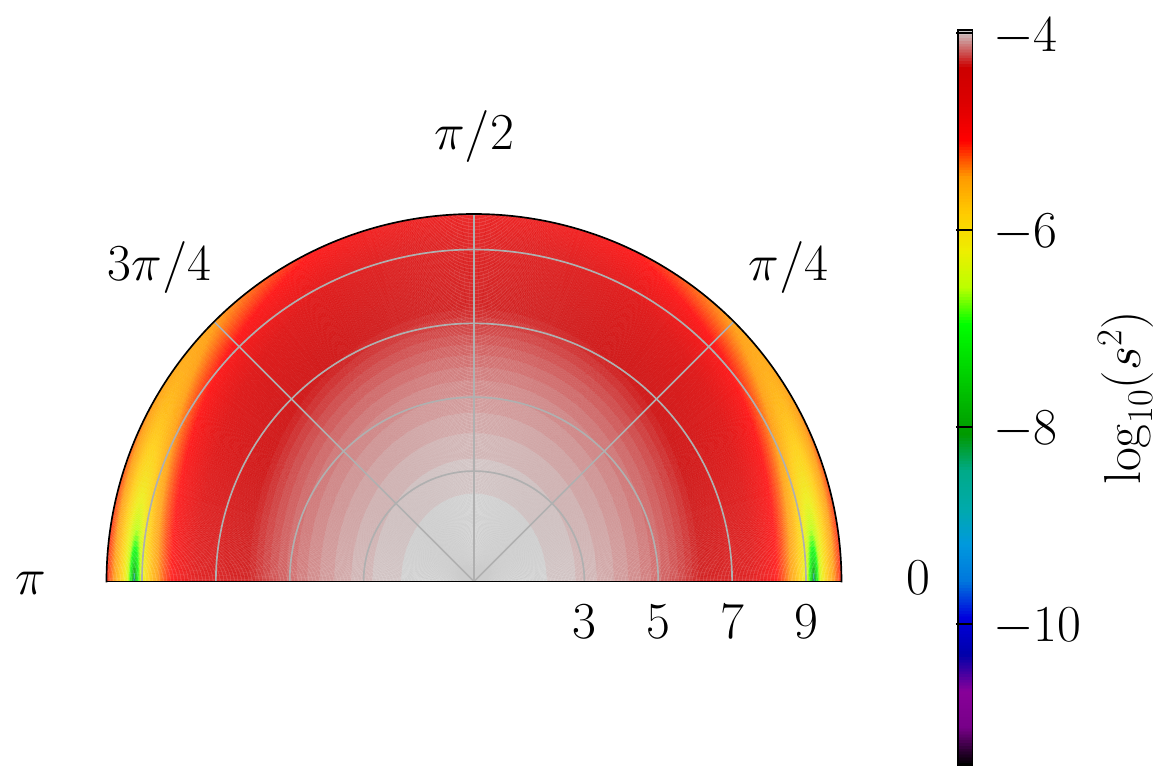}
    \caption{Distribution of the strain $s$ within the stellar interior at a GW frequency of $100\,\rm Hz$, for a $1.4\,M_{\odot}$ strangeon star with $\mu = 10^{34}\, \mathrm{erg}\,\mathrm{cm}^{-3}$. The polar axis points toward the companion star. The full three-dimensional distribution can be obtained by rotating this cross-section $180^\circ$ around the polar axis. The strain is largest near the stellar center.}
    \label{fig:strain}
\end{figure}

The corresponding radial profiles of the perturbation variables are shown in
Fig.\,\ref{fig:figure5}. Taking $f_{\rm GW} = 100\,\rm Hz$, the resulting strain
distribution is shown in Fig.\,\ref{fig:strain}.

Once the accumulated strain reaches the breaking strain $\sigma$, the star
fractures. Large-scale fracturing drastically reduces the maximum stress the
interior can sustain, so its tidal response shifts from that of a solid body
toward fluid-like behavior—a change in tidal deformability that is directly
imprinted on the gravitational waveform and its phase
evolution~\citep{Lai:2018ugk}. Locating the onset of fracturing is therefore
crucial.

As shown in Fig.\,\ref{fig:strain}, the strain at the stellar center is significantly greater than that at the surface, suggesting that fracturing is more likely to initiate in the core. This behavior, however, differs from the assumption adopted by \citet{Zhou:2023dcf}, where fracturing was assumed to begin at the surface. Qualitatively, since strain measures relative displacement, a large deformation at the surface does not necessarily imply a large strain—because the reference length scale there is also large. In contrast, near the center even a small displacement is taken up over a correspondingly small length scale, so the strain—the relative displacement—reaches its maximum at the center rather than at the surface. This behavior also appears in the Newtonian solution (see Appendix\,\ref{app:C}), providing independent confirmation.

\begin{figure*}
    \centering
    \includegraphics[width = 0.8\textwidth]{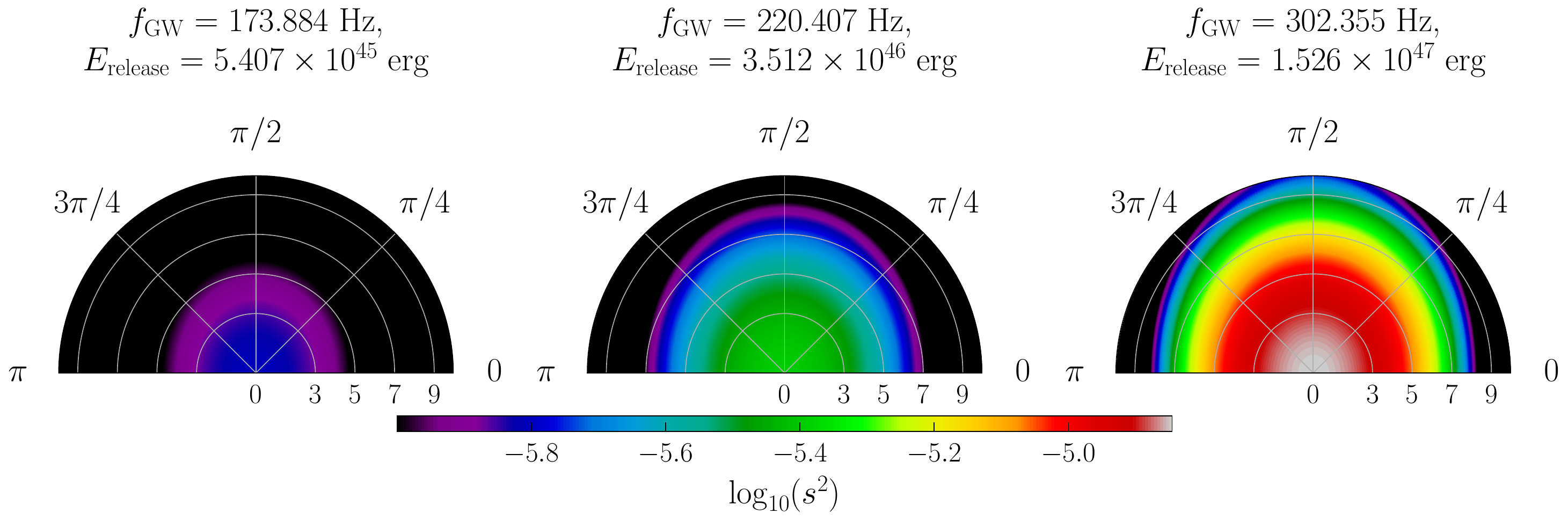}
    \caption{Progressive fracturing of a $1.4\,M_{\odot}$ strangeon star during binary inspiral, with $\mu = 10^{34}\,\rm erg\,\rm cm ^{-3}$ and breaking strain $\sigma = 0.001$. From left to right, the panels show the strain distribution at GW frequencies corresponding to 14\%, 50\%, and 86\% of the stellar volume exceeding the breaking strain. Black regions indicate intact material where $s < \sigma$.}
    \label{fig:s2frg}
\end{figure*}

We further note that the strain at the equator on the stellar surface is larger than that in the polar regions (see Fig.\,\ref{fig:s2frg}). A similar equatorial preference is also found by~\citet{Lu:2023dwi} for spin-down induced strain accumulation in SnSs, suggesting a common geometric feature of tidal and spin-down in solid compact stars. This preference, together with its subsequent breaking and recovery processes~\citep{Lai:2023axr}, could explain the observed anti-correlation between glitch amplitude and recovery fraction~\citep{Lai:2017xys}, as well as the approximately exponential recovery of the spin frequency following a glitch. In addition, starquakes with an equatorial preference that trigger coherent radio emissions in closed field lines would explain why it is so difficult to measure the spin periods of repeating fast radio bursts (FRBs)~\citep{XU:2024ycj}.

The GW frequency tracks the stage of the binary inspiral, as illustrated in
Fig.\,\ref{fig:s2frg}. As the tidal field strengthens, the strain exceeds the
breaking strain $\sigma$ over a growing internal region
$D=\{\vec{r}\,|\,s(f_{\mathrm{GW}},\vec{r})\geq\sigma\}$, which stores an
increasing amount of elastic energy. By the final stage of the inspiral, when
the two stars are separated by $20$--$30\,\rm km$, the fracture has spread across
about 80\% of the stellar volume. Assuming that upon fracturing all elastic energy within
this region is released while the remaining volume remains unaffected, an upper
bound on the released elastic energy can be estimated as:

\begin{equation}
\label{eq:eng}
    E_{\mathrm{release}}(f_{\mathrm{GW}})=\int_D\mu s^2(f_{\mathrm{GW}},\vec{r})\sqrt{-g}\, \dd V\,.
\end{equation}

\begin{figure}
    \centering
    \includegraphics[width=0.48\textwidth]{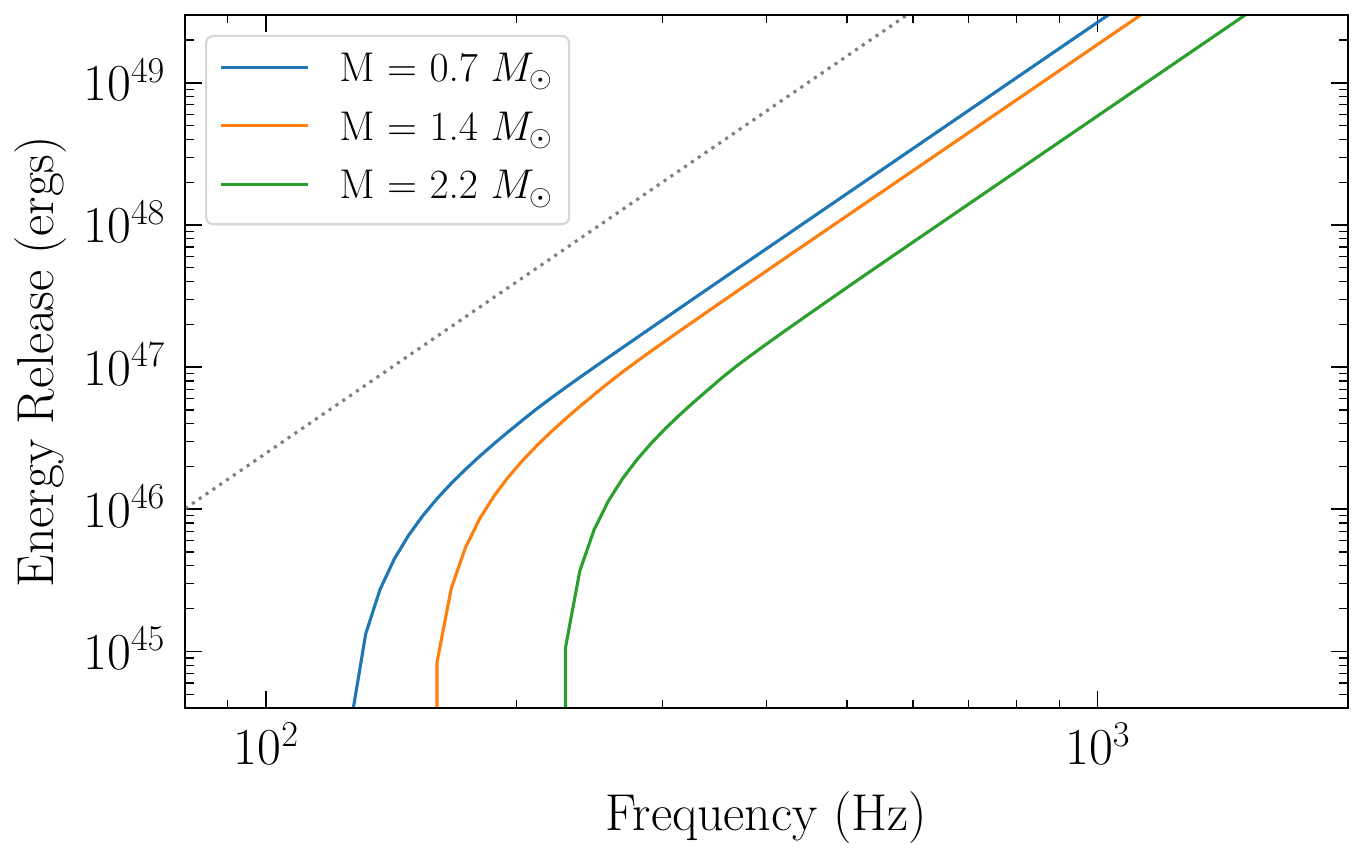}
    \caption{Released elastic energy as a function of GW frequency for strangeon stars of different masses, with a $1.4\,M_{\odot}$ companion, $\mu = 10^{34}\,\rm erg\,\rm cm ^{-3}$, and breaking strain $\sigma = 0.001$. The energy is $\sim 10^{45}$--$10^{46}\,\mathrm{erg}$ at the onset of large-scale fracturing and rises to $\sim 10^{49}\,\mathrm{erg}$ by merger. The dashed grey line shows the $E\propto f^{4}$ scaling estimated by \citet{Zhou:2023dcf}.}
    \label{fig:Energy-fre}
\end{figure}

The numerical results are presented in Fig.\,\ref{fig:Energy-fre}. For
intermediate-mass stars, the released energy is $\sim 10^{45}$--$10^{46}\,
\mathrm{erg}$ at the onset of large-scale fracturing (several hundred $\rm Hz$)
and grows as $E\propto f_{\rm GW}^4$ to $\sim 10^{49}\,\mathrm{erg}$ by merger.
This range brackets the energies of the phenomena it might power—typical short
GRB precursors ($\sim 10^{46}$--$10^{48}\,\mathrm{erg}$) and FRBs ($\sim
10^{36}$--$10^{43}\,\mathrm{erg}$)~\citep{Wang:2024opz}—supporting the solid-star
hypothesis.

\begin{figure}
    \centering
    \includegraphics[width=0.48\textwidth]{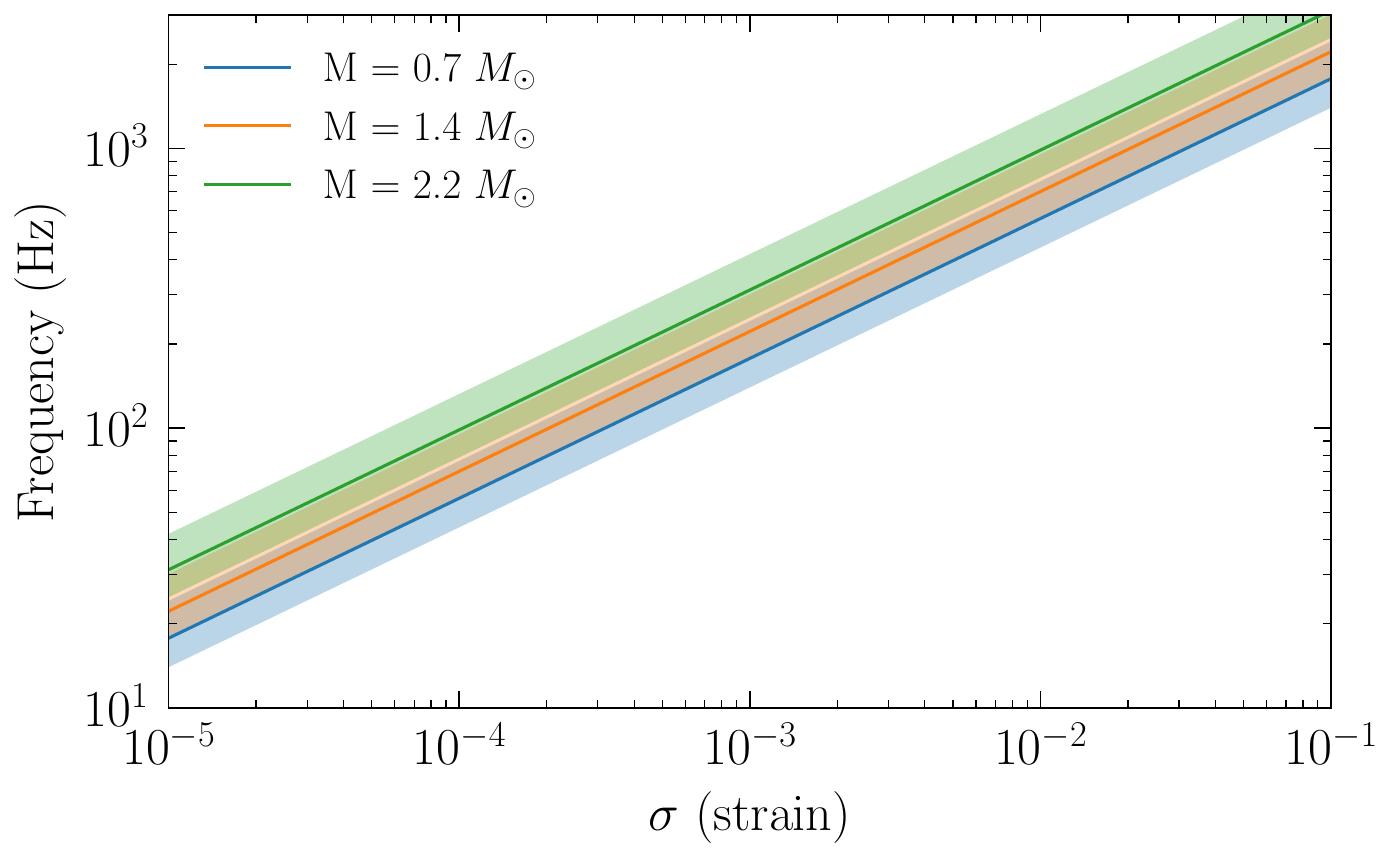}
    \caption{GW frequency at which 50\% of the stellar volume is fractured, as a function of the breaking strain $\sigma$, for strangeon stars of different masses with a $1.4\,M_{\odot}$ companion. The shaded bands span the frequency range between 14\% and 86\% fractured volume.}
    \label{fig:fre-sig}
\end{figure}

However, the breaking strain remains uncertain, with
estimates ranging from $\sim 10^{-5}$ to $\sim 0.1$~\citep{Smoluchowski:1970zz,
Caplan:2018gkr}. For strangeon matter, this value is typically assumed to be on
the order of $0.01$, although processes such as fault slip could reduce the
effective threshold. Fig.\,\ref{fig:fre-sig} shows the GW frequency at which
half of the star's volume is fractured, plotted as a function of this threshold.

\citet{Zhou:2023dcf} recently explored the possibility that the precursor
emission of GRB 211211A originates from the release of elastic energy during a
starquake. For a star of a given mass, the elastic energy released at complete fracture
scales as $E_{\rm release} \propto \mu f_{\rm GW}^4$
(Eq.\,\ref{eq:eng} and Fig.\,\ref{fig:Energy-fre}), consistent
with \citet{Zhou:2023dcf}. The frequency of complete fracture, in turn, depends
sensitively on the breaking strain $\sigma$, as illustrated in
Fig.\,\ref{fig:fre-sig}. For example, for
$\sigma=0.001$, we expect $f_{\rm GW}\sim 200 \rm Hz$, corresponding to a
separation of $\sim 50\,\rm km$, which is consistent with the observations of
GRBs with short waiting times~\citep{Suvorov:2024cff}. Consequently, combined
measurements of the waiting time and the intensity of the precursor could
provide joint constraints on the parameters $\mu$ and $\sigma$. Such constraints
would offer a valuable test of both the strangeon star hypothesis and the
associated starquake model.

\section{Conclusions}
\label{sec:conclusion}

In this paper, we have investigated the tidal deformability and internal strain
accumulation of solid strangeon stars. While previous studies of neutron stars
with thin solid crusts found negligible deviations in tidal deformability ($<
10^{-7}$)~\citep{Gittins:2020mll,Penner:2011pd}, and \citet{Gao:2021uus} showed
that the I--Love relation for fluid SnSs closely follows that of conventional
neutron stars, the present work demonstrates that accounting for the fully solid
nature of SnSs leads to qualitatively different results.

Our key findings are as follows. First, the tidal deformability of solid SnSs
differs by approximately $40\%$ from their fluid counterparts—a discrepancy
large enough to produce observable imprints in GW signals~\citep{Creci:2024wfu}
and that may be accessible with next-generation detectors. This result is
consistent with previous studies of solid SqSs~\citep{Lau:2018mae} and extends
the analysis to the strangeon star framework. Second, detailed modeling of
internal strain accumulation reveals that fracturing initiates at the stellar
center—contrary to the surface-fracturing assumption of \citet{Zhou:2023dcf}—and
that large-scale fracturing occurs within the several-hundred-Hz GW frequency
band. Third,
the elastic energy released during fracturing is $\sim 10^{45}$--$10^{46}\,\mathrm{erg}$
at onset and grows as $E\propto\mu f_{\rm GW}^4$ to $\sim 10^{49}\,\mathrm{erg}$
by merger, spanning the energies of observed GRB
precursors~\citep{Zhou:2023dcf} and FRBs~\citep{Wang:2024opz}.

Taken together, these signatures are complementary: the precursor intensity and
waiting time encode the shear modulus $\mu$; the solid–fluid difference in tidal
deformability provides an independent diagnostic; and the abrupt waveform change
at large-scale fracturing constrains both $\mu$ and $\sigma$. A multi-messenger
observation combining the GW signal with the precursor electromagnetic emission
could therefore pin down $\mu$ and $\sigma$—or, if the two tests prove mutually
inconsistent, rule out the solid-star hypothesis, and with it the strangeon-star
conjecture.

Several limitations of this work should be noted. The fracturing process and
energy release are modeled using a simplified single-layer configuration. A more
realistic treatment would require a two-layer structure with a fluid-like core
and a solid envelope. As discussed in Section~\ref{subsec:strain}, the solid
core may fracture extensively during the early inspiral and subsequently behave
as an effective fluid. Such a two-layer model has been explored for solid SqSs
by \citet{Lau:2018mae}, who showed that the screening factor gradually decreases
as the solid envelope thins. Assuming that this picture extends to SnSs, our
single-layer approximation remains valid during most of the inspiral, when the
fluid core radius is below $0.5R$~\citep{Lau:2018mae}. Additionally, the breaking
strain $\sigma$ remains poorly constrained, spanning several orders of
magnitude~\citep{Smoluchowski:1970zz, Caplan:2018gkr}. Future work should
incorporate a fully time-evolving two-layer simulation and explore how different
fracture criteria affect the observable signatures.

\section*{Acknowledgements}
This work was supported by the National Natural Science Foundation of China
(12447148), the China Postdoctoral Science Foundation (2024M760081), the
National SKA Program of China (2020SKA0120100), and the High-Performance
Computing Platform of Peking University.

\section*{Data Availability}
The data underlying this article will be shared on reasonable request to the
corresponding author.

\bibliography{apssamp}

\onecolumn
\renewenvironment{widetext}{}{}

\appendix
\section{Perturbation equations}
\label{app:A}

Define $L_1=l\,(l+1)$. In subsequent calculations, the case $l=2,m=0$ is
adopted. Defining the squared sound speed as $c_s^2=\frac{\dd P}{\dd \rho}$, we
introduce the following coefficients:
\begin{align}
\alpha_1& = \mu\, ,\\
\alpha_2& = c_s^2(\rho+P)-\frac{2}{3}\mu\, ,\\
\alpha_3& = c_s^2(\rho+P)+\frac{4}{3}\mu\, .
\end{align}

Simplifying the perturbation equations, we obtain:
\begin{widetext}
\begin{equation}
\label{eq:dif1}
\frac{\dd W}{\dd r} = \left(1 - \frac{2\alpha_2}{\alpha_3} - \frac{r \lambda'}{2}\right)\frac{W}{r} - \frac{r Z_r}{\alpha_3} 
               + \frac{\alpha_2}{\alpha_3} \frac{L_1 V}{r} - \frac{\alpha_2}{\alpha_3} r K - \frac{1}{2} r H_2\, , 
\end{equation}

\begin{equation}
\frac{\dd V}{\dd r} = -e^\lambda \frac{W}{r} + 2\frac{V}{r} - r e^\lambda \frac{Z_{\perp}}{\alpha_1}\, ,
\end{equation}

\begin{equation}
\frac{\dd H_0}{\dd r} = J + \frac{(\nu' + \lambda') W}{r^2} - 16\pi \alpha_1 \nu' V\, ,
\end{equation}
\end{widetext}

\begin{widetext}
\begin{equation}
\begin{split}
\frac{\dd J}{\dd r} &= \left[ \frac{32\pi e^\lambda}{r^2} \alpha_1\left(1 + \frac{2\alpha_2}{\alpha_3}\right) - \frac{3 \nu' (\lambda' + \nu')}{2r^2} \right] W
- 8\pi e^\lambda \left(1 + \frac{2\alpha_2}{\alpha_3}\right) Z_r \\
&\quad - \frac{8\pi}{r^2}\Bigl[ (\rho + P) e^\lambda L_1 
 + 2\alpha_1\left(1 + \frac{2\alpha_2}{\alpha_3}\right) e^\lambda L_1 
  + 4\alpha_1 (1 - e^\lambda) - 2\alpha_1 (r \nu')^2 \Bigr] V \\
 &\quad - 16\pi e^\lambda (r \nu') Z_{\perp} 
  + \left[ L_1 e^\lambda + 2(e^\lambda - 1) - r\left(\frac{\lambda'}{2} + \frac{5\nu'}{2}\right) 
  + (r \nu')^2 \right] \frac{H_0}{r^2} \\
  &\quad + \left[ \frac{r(\lambda' - \nu')}{2} - 2 \right] \frac{J}{r} 
  + \left[ \frac{\lambda' + \nu'}{r} 
  + 16\pi e^\lambda \alpha_1\left(1 + \frac{2\alpha_2}{\alpha_3}\right) \right] K\, ,
\end{split}
\end{equation}
\end{widetext}

\begin{widetext}
\begin{equation}
    \begin{split}
\frac{\dd Z_r}{\dd r} &= \left[ P'\left( \frac{r \nu''}{\nu'} - \frac{r \lambda'}{2} - 2 \right)
    - \frac{4\alpha_1}{r}\left(1 + \frac{2\alpha_2}{\alpha_3}\right) \right] \frac{W}{r^2} 
    - \left( \frac{r \nu'}{2} + \frac{4\alpha_1}{\alpha_3} \right) \frac{Z_r}{r} \\
    &\quad + \left( P' + \frac{2\alpha_1}{r}\left(1 + \frac{2\alpha_2}{\alpha_3}\right) \right) \frac{L_1 V}{r^2} 
    + e^\lambda L_1 \frac{Z_{\perp}}{r} + \frac{(\rho + P)}{2} \frac{\dd H_0}{\dd r}\quad - P' \frac{H_2}{2} 
    - \left[ P' + \frac{2\alpha_1}{r}\left(1 + \frac{2\alpha_2}{\alpha_3}\right) \right] K\, ,
    \end{split}
\end{equation}

\begin{equation}
\label{eq:dif2}
\begin{split}
\frac{\dd Z_{\perp}}{\dd r} &= \left[ P' + \frac{2\alpha_1}{r}\left(1 + \frac{2\alpha_2}{\alpha_3}\right) \right] \frac{W}{r^2} 
- \frac{\alpha_2}{\alpha_3} \frac{Z_r}{r}\quad - \left[ -\frac{2\alpha_1}{r} + \frac{2\alpha_1}{r}\left(1 + \frac{\alpha_2}{\alpha_3}\right) L_1 \right] \frac{V}{r^2} 
- \left( \frac{r \lambda'}{2} + \frac{r \nu'}{2} + 3 \right) \frac{Z_{\perp}}{r} \\
&\quad + \frac{(\rho + P)}{2} \frac{H_0}{r}
+ \frac{\alpha_1}{r}\left(1 + \frac{2\alpha_2}{\alpha_3}\right) K\, .
\end{split}
\end{equation}
\end{widetext}

Additionally, the system involves the algebraic relations:
\begin{equation}
H_2=H_0+32\pi\alpha_1 V\, ,
\end{equation}
\begin{equation}
        (L_1-2)e^{\lambda}K=[(\nu')^2+\nu'\lambda'+16\pi e^{\lambda}rP']W
        -16\pi e^{\lambda}r^2Z_r-16\pi e^{\lambda}(2+r\nu')r^2 Z_{\perp}
        +[L_1e^{\lambda}-2+(r\nu')^2]H_0+r^2\nu'J\, .
\end{equation}

Thus, the resulting system of six first-order differential equations is
complete. We have verified this system and confirmed through calculation that it
is equivalent to the equations presented by~\citet{Gao:2025aqo}.

\section{Boundary condition: center}
\label{app:B}

The Taylor expansions of the background solution parameters around the center
are given by:
\begin{align}
    \rho &= \rho_c+r^2\rho_2\, ,\\
    P &= P_c+r^2P_2\, .
\end{align}

Requiring that the coefficients of the divergent terms vanish and that the
leading non-zero terms match order by order, the eigensolutions are as follows:
\begin{widetext}
\begin{equation}
    W^{(0)} = 2 V^{(0)}\, ,
\end{equation}
\begin{equation}
\begin{split}
    W^{(2)} &= \frac{1}{\gamma P_c \cdot 5 + \mu \cdot \dfrac{11}{3}}\Bigg[- \bigl( \rho_c + (3\gamma + 1) P_c \bigr) \frac{H_{0}^{(0)}}{2} + \frac{4}{9} \Bigl(18\pi \mu P_c (1 - 6\gamma)\\
    +& 18\pi P_c (\rho_c + P_c - 2\mu) + \pi \rho_c \bigl(6 (\rho_c + (1 - 2\gamma) P_c) - 10\mu\bigr)\Bigr) V^{(0)} + 3 \bigl( \gamma P_c \cdot 2 - \frac{4\mu}{3} \bigr) V^{(2)}    \Bigg]\, ,
\end{split}
\end{equation}
\begin{equation}
\begin{split}
    H_{0}^{(2)} &= \frac{1}{14} \Bigg[ 
            -4\pi \bigl( \frac{\rho_c}{3} + 3(5 - 9\gamma) P_c \bigr) H_{0}^{(0)} - \frac{32\pi^2}{3} \Bigl( 
                    6 P_c \bigl( (3 + \frac{1}{c_s^2})(\rho_c + P_c) - 4\mu \bigr) \\
            &\qquad + 2\rho_c \bigl( 
                        (1 + \frac{1}{c_s^2})(\rho_c + P_c) - 6\gamma P_c - \mu 
                    \bigr) + 12\mu (-\rho_c + (3 - 9\gamma) P_c) 
                \Bigr) V^{(0)} \\
            &\quad - 8\pi \bigl( \rho_c + (1 + 3\gamma) P_c \bigr) (-5 W^{(2)} + 6 V^{(2)}) 
            \Bigg]\, ,
\end{split}
\end{equation}
\begin{equation}
    Z_{\perp}^{(0)} = -2\mu V^{(0)}\, ,
\end{equation}
\begin{equation} 
    Z_{\perp}^{(2)} = -\mu (W^{(2)} + 2 V^{(2)})\, ,
\end{equation}
\begin{equation}
    Z_{r}^{(0)} = -4\mu V^{(0)}\, ,
\end{equation}
\end{widetext}

\begin{widetext}
\begin{equation}
\begin{split}
    Z_{r}^{(2)} &= -\Bigl[ \frac{16\pi}{3} \rho_c \bigl( c_s^2 (\rho_c + P_c) + \frac{4}{3}\mu \bigr) + 16\pi \cdot 3 c_s^2 (\rho_c + P_c) \mu \Bigr] V^{(0)} \\
     &\quad - \frac{1}{2} \cdot 3 c_s^2 (\rho_c + P_c) H_{0}^{(0)} - \bigl( 5 c_s^2 (\rho_c + P_c) + \frac{8}{3}\mu \bigr) W^{(2)} + 6 \bigl( c_s^2 (\rho_c + P_c) - \frac{2}{3}\mu \bigr) V^{(2)}\, ,
\end{split}
\end{equation}
\begin{equation}
    J^{(0)} = 2 H_{0}^{(0)} - 8\pi (\rho_c + P_c) V^{(0)}\, ,
\end{equation}
\begin{equation}
    J^{(2)} = 4 H_{0}^{(2)} 
            - \frac{8\pi}{3} \Bigl( 
                16\pi \mu (\rho_c + 3 P_c) 
                + 6 \bigl( \frac{P_2}{c_s^2} + P_2 \bigr) 
            \Bigr) V^{(0)} + 8\pi (\rho_c + P_c) W^{(2)}\, ,
\end{equation}
\end{widetext}
where
\begin{align}
    \gamma =& \frac{P_c+\rho_c}{P_c}\left(\frac{\dd P}{\dd \rho}\right)_{r=0}\, ,\\
    P_2 =& -\frac{4\pi}{3}(\rho_c+3P_c)(\rho_c+P_c)\, ,\\
    \rho_2 =& \frac{P_2}{c_s^2}\, .
\end{align}

This result shows that there are only three independent initial conditions at the stellar center. Together with the two boundary equations at the surface, the solution is determined up to a single overall scaling factor (see Section\,\ref{subsec:Perturb}).

\section{Solutions under Newtonian gravity}
\label{app:C}

Within the Newtonian framework, the tidal field can be described as:
\begin{equation}
    \psi_{\rm tidal} =  r^2\mathcal{E}\,Y_l^m(\theta,\phi)\,.
\end{equation}
For the tidal component, we take $l = 2$ and $m = 0$. The displacement can then be expanded in spherical harmonics, following the similar procedure used in the general-relativistic treatment (see Section\,\ref{subsec:Perturb}):

\begin{align}
    \label{eq:appc0u}
    u_r &= U(r) Y_l^m(\theta,\phi)\,,\\
    u_{\theta} &= V(r) \partial_{\theta}Y_l^m(\theta,\phi)\,,\\
    \label{eq:appc0d}
    u_{\phi} &= \frac{V(r)}{\sin^2\theta}\partial_{\phi}Y_l^m(\theta,\phi)\,.
\end{align}

With the displacement known, the strain tensor and stress tensor are given by:
\begin{equation}
    s_{ab} = (\nabla_a u_b+\nabla_b u_a)/2\,,
\end{equation}
\begin{equation}
    \sigma_{ab} = -p g_{ab} + 2\mu s_{ab}\,.
\end{equation}

For a static equilibrium state, the gravitational potential $\psi$ and the stress tensor must satisfy the Poisson equation and the force balance equation:
\begin{align}
    \label{eq:app_possionu}
    \nabla^2 \psi &= -4\pi G \rho\,,\\
    \label{eq:app_possiond}
    0 &= \nabla^b \sigma_{ba} - \rho \nabla_a \psi\,.
\end{align}

Here $\psi$ denotes the gravitational potential and $\rho$ the mass density. In the presence of a tidal field, each quantity is decomposed into an unperturbed part and a perturbation:
\begin{align}
    \psi &= \psi_0 +\psi_1\,,\\
    p &= p_0 +\delta p\,,\\
    \rho &= \rho_0 +\delta \rho\,.
\end{align}
with the unperturbed displacement field vanishing ($u_a = 0$). Substituting these expansions into Eq.\,(\ref{eq:app_possionu}--\ref{eq:app_possiond}) keeping only first-order terms, we obtain the linearized perturbation equations~\citep{Greff-Lefftz2005}:
\begin{gather}
    \label{eq:appc1u}
    \nabla^2 \psi_1 = -4\pi G \delta\rho\,,\\
    \label{eq:appc1d}
    0 = -\nabla_a \delta p +2\mu \nabla^b s_{ba} - \rho_0 \nabla_a \psi_1\,.
\end{gather}

For a compressible fluid, the pressure perturbation is related to the density perturbation by $\delta p = \kappa\, \delta \rho$, where $\kappa$ is the bulk modulus. In the incompressible limit $\kappa\rightarrow\infty$, a finite $\delta p$ requires $\delta\rho\rightarrow 0$; equivalently, mass conservation $\delta\rho = -\rho_0\nabla^a u_a$ then forces the displacement field to be divergence-free,
\begin{equation}
    \label{eq:appc2}
    \nabla^a u_a = 0\,.
\end{equation}

Substituting Eq.\,(\ref{eq:appc2}) into the perturbation equations, we obtain the governing equations for the incompressible problem:
\begin{gather}
    \label{eq:appc_perturbu}
    \nabla^2 \psi_1 = 0\,,\\
    \nabla^a u_a = 0\,,\\
    \label{eq:appc_perturbd}
    0 = -\nabla_a \delta p +2\mu \nabla^b s_{ba} - \rho_0 \nabla_a \psi_1\,.
\end{gather}

Expanding the perturbations in spherical harmonics, $\psi_1 = \psi_1(r)\,Y_l^m(\theta,\phi),\,\delta p = \delta p(r)\,Y_l^m(\theta,\phi)$, and specializing to the tidal component $l=2$, $m=0$, we obtain from Eqs.\,(\ref{eq:appc_perturbu}-\ref{eq:appc_perturbd}) the following system of ordinary differential equations:
\begin{align}
    \label{eq:appc31}
    -6\psi_1(r)&+r(2\psi'_1(r)+r\psi''_1(r)) = 0\,,\\
    2rU(r)-&6V(r)+r^2U'(r)=0\,,\\
    -10r\mu U(r)&+24\mu V(r) -r(-4r\mu U'(r)+6\mu V'(r)+r^2(\delta p'(r)+\rho_0 \psi'_1(r)))+2r^3 \mu U''(r)=0\,,\\
    -4r\mu U(r)&+12\mu V(r)+r^2(\delta p(r)+\rho_0 \psi_1(r)-\mu (U'(r)+V''(r)))=0\,.
\end{align}

Thus we have four equations for the four unknown radial functions $U(r)$, $V(r)$, $\psi_1(r)$, and $\delta p(r)$. This system can be solved once appropriate boundary conditions are imposed. We first require regularity at the center, i.e., all quantities remain finite as $r \to 0$. For a star of constant density $\rho_0$, the analytical solution takes the form
\begin{gather}
    U(r) = c_1 r +c_2 r^3\,,\\
    V(r) = \frac{1}{6}(3c_1 r^2 + 5 c_2 r^4)\,,\\
    \psi_1(r) = c_3 r^2\,,\\
    \delta p = (7c_2\mu - c_3 \rho_0)r^2\,.    
\end{gather}

At the stellar surface ($r = R$), the total force must vanish, which imposes the following boundary conditions~\citep{Love1911,Greff-Lefftz2005}:
\begin{gather}
    \sigma_{rr} + \rho_0 g u_r = 0\,,\\
    \sigma_{r\theta} = 0\,,
\end{gather}
where $g$ is the surface gravitational acceleration. Applying these conditions to the general solution above determines the constants $c_2$ and $c_3$ in terms of $c_1$, leaving the overall solution determined up to a single scaling factor $c_1$:

\begin{gather}
\label{eq:appc4}
    U(r) = c_1 (r - \frac{3}{8} \frac{r^3}{R^2})\,,\\
    V(r) = \frac{c_1}{16}r^2(8 - 5\frac{r^2}{R^2})\,,\\
    \psi_1(r) = -c_1 r^2 \frac{19\mu+5gR\rho_0 }{8R^2\rho_0}\,,\\
    \delta p = c_1 r^2 \frac{-2\mu+5gR\rho_0}{8R^2}\,,
\end{gather}
where $R$ is the star's radius. 

Having obtained the interior solution, we now match it to the external tidal field at the stellar surface $r = R$. At the surface, the density jumps discontinuously from $\rho_0$ inside to $0$ outside. According to Eqs.~(\ref{eq:appc1u}--\ref{eq:appc1d}), this density discontinuity leads to a jump in the radial derivative of the gravitational potential~\citep{Greff-Lefftz2005}:
\begin{equation}
\label{eq:appc_jump}
\psi'(R^+) - \psi'(R^-) = -4\pi G U(R) (0 - \rho_0) = 4\pi G \rho_0 U(R)\, .
\end{equation}

The total gravitational potential consists of two parts: the external tidal potential $\mathcal{E}$ and the star's self-gravitational response $Q$. For the dominant $l=2$, $m=0$ tidal component, we write
\begin{equation}
    \psi_1(r) =
    \begin{cases}
    -\frac{3}{2}Q r^2/R^5 + \frac{1}{2}\mathcal{E}r^2\;\; r<R\,;\\
    -\frac{3}{2}Q/r^3 + \frac{1}{2}\mathcal{E}r^2\;\; r>R\,.
    \end{cases}
\end{equation}

The continuity of the potential at $r = R$ gives
\begin{equation}
\psi_1(R) = -\frac{3}{2}\frac{Q}{R^3} + \frac{1}{2}\mathcal{E} R^2.
\end{equation}

Applying the jump condition Eq.~\eqref{eq:appc_jump} and the potential continuity, the constants $c_1$, $Q$, and $\mathcal{E}$ can be related. Solving the matching conditions yields the explicit expression~\citep{1960regd.book.....M}:
\begin{equation}
    k_2 = - \frac{3Q}{2\mathcal{E} R^5} = \frac{3/4}{1+\frac{19\mu}{2g\rho_0 R}} = \frac{3/4}{1+\tilde{\mu}}\,,
\end{equation}
where $\tilde{\mu} \equiv 19\mu/(2g\rho_0 R)$. This result confirms the Newtonian limit quoted in Eq.\,(\ref{eq:k2newton}) of the main text.

Finally, substituting the solutions for $U(r)$, $V(r)$, and $\psi_1(r)$ into the definition of the strain invariant $s^2 = \frac{3}{2}s_{ab}s^{ab}$ (see Eq.\,(\ref{eq:s2}) in the main text), we obtain the spatial distribution of the strain inside the star:
\begin{equation}
\label{eq:appcs2}
    s^2 = \frac{3}{2}s_{ab}s^{ab}=\frac{45}{2}c_1^2(471r^4-960r^2R^2+512R^4
    +24r^2(7r^2-8R^2)\cos(2\theta)+9r^4\cos(4\theta))/(4096\pi R^4)\,.
\end{equation}
This expression shows that the strain is enhanced towards 
the center, and the constant $c_1$ is determined by the tidal 
field strength through the matching conditions discussed above.
Such distribution is consistent with the results obtained in the 
general-relativistic treatment (see Fig.\,\ref{fig:strain} in the main text).

\bsp
\label{lastpage}
\end{document}